\documentclass[%
reprint,
 amsmath,amssymb,
 aps,
]{revtex4-1}

\usepackage{graphicx}
\usepackage{dcolumn}
\usepackage{bm}
\usepackage{epstopdf}
\usepackage{color}
\usepackage{amsmath}
\DeclareMathOperator{\sech}{sech}

\begin{document}
\preprint{APS/123-QED}

\title{ Quantum metrology  beyond Heisenberg limit with entangled matter wave solitons}

\author{D. V. Tsarev}
\affiliation{National Research University for Information Technology,
Mechanics and Optics (ITMO), St. Petersburg 197101, Russia}

\author{S. M. Arakelian}
\affiliation{Vladimir State University named after A. G. and N. G. Stoletovs, Gorkii Street 87, Vladimir, Russia}

\author{You-Lin Chuang}
\affiliation{Physics Division, National Center for Theoretical Sciences, Hsinchu 30013, Taiwan}

\author{Ray-Kuang Lee}
\affiliation{Physics Division, National Center for Theoretical Sciences, Hsinchu 30013, Taiwan}
\affiliation{Institute of Photonics Technologies, National Tsing Hua University, Hsinchu 30013, Taiwan}
\affiliation{Department of Physics, National Tsing Hua University, Hsinchu 30013, Taiwan}

\author{A. P. Alodjants}
\affiliation{National Research University for Information Technology,
Mechanics and Optics (ITMO), St. Petersburg 197101, Russia}
\affiliation{Vladimir State University named after A. G. and N. G. Stoletovs, Gorkii Street 87, Vladimir, Russia}

\date{\today}

\date{\today}

\begin{abstract}
By considering matter wave bright solitons from weakly coupled Bose-Einstein condensates trapped in a double-well potential, we study the formation of macroscopic non-classical states, including Schr\"odinger-cat  superposition states and maximally path entangled $N00N$-states.
With these macroscopic states, we examine Mach-Zehnder interferometer in the context of parity measurements, in order to obtain Heisenberg limit accuracy for linear phase shift measurement.
We reveal that the ratio between two-body scattering length and intra-well hopping parameter can be measured with the scaling beyond this limit by using nonlinear phase shift with interacting quantum solitons.
\end{abstract}

\maketitle

\section{Introduction}
Nowadays quantum metrology has become one of fascinating areas in modern quantum physics,  dealing with new approaches for measurement, control, and estimation of physical parameters to achieve limiting accuracy and to explore all facilities of current quantum technologies~\cite{wiseman,wieman,dowling15,giovannetti,roy,boixo}.
Apart from classical measurement theory,  quantum approach predicts so-called the quantum Cramer-Rao (QCR) bound  $\langle(\delta \phi_{est})^2\rangle_\phi\geq[\nu F(\phi)]^{-1}$ for estimating arbitrary physical parameter $\phi$ within a set of $\nu$ trials through Fisher information $F(\phi)$~\cite{dowling15}.
In particular,  phase estimation requires high precision measurement, which can be realized both in optical ~\cite{Caves81, yurke} or atomic systems ~\cite{gustavson,gross}.
The existence of standard quantum limit (SQL) sets a constraint on the linear phase shift ($\phi$) measured with an error  $\sigma_{\phi}\sim N^{-1/2}$. Here, $N$ is the average number of particles.
To break the classical limit, non-classical squeezed states have been illustrated to provide possible quantum approaches to measurement theory beyond the SQL ~\cite{Caves81, yurke}.

Surpassing SQL in the phase measurement has been demonstrated experimentally  with two-mode systems, such as Mach-Zehnder interferometers (MZI), gyroscopes, and lithography devises,  where non-classical squeezed or correlated states are applied  as the input states ~\cite{dowling98,winland,boto}.
For the linear phase measurement,  one can achieve the {\it Heisenberg limit} with the accuracy
\begin{equation}\label{acc}
\sigma_{\phi} \geq \ N^{-1},
\end{equation}
which gives the limiting case on  QCR bound related to the single mode  passing ~\cite{giovannetti}.
With the maximally entangled $N$-particle state, coined as the NOON-state, it is proven that
 for arbitrary two-mode quantum interferometers one can saturate the Heisenberg limit shown in Eq.~\eqref{acc} for quantum metrology ~\cite{boto,dowling08,pezze}.
Nevertheless, the preparation and formation of $N00N$-states,  with a large number of particles,  represents a great challenge and nontrivial task  both in theory and experiment for all the possible development in  quantum technologies~\cite{afek,rozema}.
Scaling beyond Heisenberg limit, referred as  super-Heisenberg scaling,  can be achieved in the framework of interaction-based (nonlinear) quantum metrology~\cite{roy,boixo, Luis, napolitano, Mundo}.

In this work,  we propose an alternative method to create $N00N$-states,  which are maximally entangled states in path, by means of matter wave bright solitons  in Bose-Einstein condensates (BECs).
Starting with  Gross-Pitaevskii equation (GPE) for a condensate in a double-well potential~\cite{pethick}, we describe the corresponding quantum field model for coupled bright solitons occurring in  two trapped condensates.
In the framework of variational approach, we derive the equations of motion for the  condensate's parameters,  i.e., the  relative phase and population imbalance between two solitons.
Then, we  show that the ground state of the system can be a quantum superposition state, forming  Schr\"odinger-cat or $N00N$-state.
Utilization of these states is revealed for quantum metrology, with focus on linear phase shift measurements.
The saturation of the linear Heisenberg limit is demonstrated for nonlinear parameter estimations.
Our results provide possible quantum metrology beyond linear Heisenberg limit with entangled matter wave solitons.


\section{Model for coupled quantum matter bright solitons}
Let us consider two  BECs, together consisting of $N$ particles, trapped in a double-well potential and weakly coupled to each other due to the Josephson effect.
This model has been applied for the studies on quantum squeezing, entanglement, and related  metrology  applications for continuous variables within the tight binding approximation~ \cite{cirac, Soren, Fu, Maz, He, Pezze}.
Experimentally, such an atomic system can be implemented with the help of highly asymmetric potentials, i.e., a cigar-shaped potential~\cite{morsch}.
Without loss of generality, the spatial distribution for the condensates are denoted along the $z$-direction.
In addition to atomic systems, exciton-polariton condensates in the microcavity is also a possible platform for our model~\cite{Hof}.

The total Hamiltonian $\hat{H}$ for BECs in a double-well potential  can be described by 
\begin{subequations}\label{ham}
\begin{equation}
\hat{H}=\hat{H}_1+\hat{H}_2+\hat{H}_{int},
\end{equation}
where $\hat{H}_{j}$  ($j=1,2$)  is the Hamiltonian for condensate particles in $j$-th  well; while $\hat{H}_{int}$ accounts  for the inter-well coupling between two sites.
In the second quantization  form, explicitly, we have
\begin{eqnarray}
&&\hat{H}_{j} = \int dz \hat{a}_{j}(z)^{\dag}\left(-\frac{1}{2M}\frac{\partial^2}{\partial z^2}  + \
\frac{U}{2}\hat{a}_{j}(z)^{\dag}\hat{a}_{j}(z)\right)\hat{a}_{j}(z)\nonumber\\
&&\hat{H}_{int} = \kappa\int dz \hat{a}_{2}(z)^{\dag}\hat{a}_{1}(z)+H.C.
\end{eqnarray}
\end{subequations}
Here,  the parameter $U$ characterizes two-body interactions, $M=sgn[m_{\text{eff}}]=\pm1$ is used as the normalized  effective particle  mass, and $\kappa$ denotes the inter-well tunneling rate.
The corresponding annihilation  (creation) operators of bosonic fields are denoted as $\hat{a}_{j}$ (${\hat{a}^\dag_{j}}$) with $j=1,2$, and obey the commutation relations:
\begin{equation}
[\hat{a}_i(z), {\hat{a}^\dag_{j}(z')}] = \delta (z-z')\,\delta_{ij}; \quad i,j =1,2.
\label{commutator}
\end{equation}

For Hamiltonian~\eqref{ham}, we suppose  that the ground state of this bosonic system is the product of $N$ single particle states~\cite{cirac}.
Physically,  this assumption is valid for BECs in the equilibrium states  at zero temperature. Thus, the collective ground state for the whole system can be written as:
\begin{equation}\label{vector}
|\Psi\rangle_N = \frac{1}{\sqrt{N!}}\left[\int^{\infty}_{-\infty}dz\left(\Psi_1\hat{a}_1^\dag + \Psi_2\hat{a}_2^\dag \right)\right]^N |0\rangle,
\end{equation}
with $|0\rangle\equiv|0\rangle_1 |0\rangle_2$ being the  two-mode vacuum state.
It is noted that the state vector shown in Eq.~\eqref{vector} relates to the Hartree approach for bosonic systems~\cite{Alod}, which is valid for a large number of particles $N$.
If we apply the variational approach based on the ansatz  $\Psi_1\equiv\Psi_1(z,t)$ and
$\Psi_2\equiv\Psi_2(z,t)$, with the unknown $z$-dependent wave-functions, one can have the corresponding Lagrangian density in the form~\cite{raghavan}:
\begin{eqnarray}\label{lagrange}
L&=&\
\sum_{j=1}^{2} \left( \frac{i}{2} \left[ \Psi_j^* \dot{\Psi}_j - \dot{\Psi}_j^* \Psi_j \right] + \frac{1}{2M}\Psi_j^*\frac{\partial^2 \Psi_j}{\partial z^2} - \frac{U}{2} \left| \Psi_j \right|^4 \right)  \nonumber\\
&-&\kappa \left( \Psi_1^* \Psi_2 + \Psi_1 \Psi_2^* \right).
\end{eqnarray}

In the limit of vanishing coupling constant $\kappa = 0$, Eq.~\eqref{lagrange} leads to the well-known GPE, which supports bright soliton solution when $MU<0$, i.e.,
\begin{equation}\label{sech}
\Psi_j = \frac{N_j}{2} \sqrt{|U|} \sech{\left(\frac{N_j |U|}{2} z \right)} e^{iM\theta_j}.
\end{equation}
Below,  we take the soliton solutions given in Eq.~\eqref{sech}  as our variational ansatz, but imposing time dependent parameters for $N_j$ and $\theta_j$  when the weakly coupling between the condensates is nonzero.
Then, we can obtain the effective Lagrangian by integrating the Lagrangian density~\eqref{lagrange}:
\begin{eqnarray}\label{lagrangian}
\L =&& \int_{-\infty}^{\infty}Ldz = -M\left(N_1 \dot{\theta}_1 + N_2 \dot{\theta}_2\right) \\
&+& \frac{U^2}{24M} \left(N_1^3 + N_2^3 \right)-\frac{4\kappa N_1 N_2}{N} I(p) \cos\left[\theta\right].\nonumber
\end{eqnarray}
Here, we have defined $p=(N_2 - N_1)/N$  and $\theta = \theta_2 - \theta_1$ as the population imbalance and phase difference, respectively.
The total number of particles is denoted by  $N = N_1 + N_2$.
Moreover, we also introduce
\begin{equation}
I(p) = \int_{0}^{\infty} \frac{dz'}{\cosh^2{(z')} + \sinh^2{(z' p)}},
\end{equation}
by applying the parabolic approximation, i.e., $I(p)\approx 1 - \alpha p^2$ with $\alpha = 0.21$.
Based on Eq.~\eqref{lagrangian}, we can go one step further by deriving the equation of motions for the population imbalance and phase difference, i.e.,  $p$  and $\theta$,
\begin{subequations}\label{basicEqs}
\begin{eqnarray}
&& \dot{p} = -\frac{1}{M}\left(1-p^2 \right) \left(1-\alpha\, p^2 \right) \sin{[\theta]},\\
&& \dot{\theta} = \Lambda p + \frac{2p}{M}\cos{[\theta]} \left[1 + \alpha -2 \alpha\, p^2 \right].
\end{eqnarray}
\end{subequations}
Here, the dots denote the derivative with respect to the dimensionless time $t'=2|\kappa|t$.
In Eqs.~\eqref{basicEqs}, a dimensionless parameter  $\Lambda =\frac{U^2 N^2}{16|\kappa|}$  is also introduced, which characterizes various regimes for  soliton interaction.

Two sets of nontrivial stationary solutions can be found for Eqs.~\eqref{basicEqs}. For the first set, we have
\begin{subequations}\label{sol1}
\begin{eqnarray}
&& p_0^2 = \frac{1}{2 \alpha} \left[1 + \alpha - \frac{\Lambda}{2} \right], \\
&&\cos(\theta_0) = -M;
\end{eqnarray}
\end{subequations}
and for the second set, we have
\begin{subequations}\label{sol2}
\begin{eqnarray}
&& p_0^2 = 1,\\
&& \cos{\theta_0} = -\frac{M\Lambda}{2(1-\alpha)}.
\end{eqnarray}
\end{subequations}

First set of nontrivial solutions given in Eqs.~\eqref{sol1} is similar to the one obtained under two-mode approximation, or equivalently the tight binding model~\cite{cirac}.
However, vital parameter of the system $\Lambda$ that we introduced  above  is proportional to $N^2$ instead of $N$ which occurs in two-mode limit, cf.~\cite{Pezze}. This fact seems to be very important in practice when we consider  limit of large particle number $N$, cf. ~\cite{Maz}.
In the following, we show that this set of solutions can be used to construct Schr\"odinger-cat state with solitons.

As for the second set of solutions given in Eqs.~\eqref{sol2}, there is no  analogy from the results obtained under  two-mode approximation~\cite{cirac, Soren, Fu, Maz, He}.
Physically, such a set of solutions implies the formation of  $N00N$-states from  coupled solitons.

As for the imbalance parameter $0\leq|p|\leq1$, the corresponding  $\Lambda$ parameter  lies between $2(1-\alpha)$ and $2(1+\alpha)$, resulting in the first set of solutions only existing in  $1.58\leq\Lambda\leq2.42$.
However, for the phase difference $0\leq|\cos{\theta_0}|\leq1$, the second set of solutions can exist for $0\leq\Lambda\leq1.58$.
One can see that there is a critical value for $\Lambda_{cr}=2(1-\alpha)\approx1.58$, at which we have a state with $p^2=1$ and  $\cos(\theta_0)=-M$.
This state at the critical value of $\Lambda_{cr}$ corresponds to a maximal population imbalance for out-of-phase solitons,  and for in-phase gap solitons.

To be more specific thereafter we assume $M=1$ that corresponds to  bright  solitons with attractive condensate particles (atoms), cf. \cite{strecker}.



\section{Superposition states of quantum solitons}
\subsection{Schr\"odinger-cat states (SCS)}

The wavefunction of solitons corresponding to Eqs.~\eqref{sol1} has the form
\begin{subequations}\label{SCS_wf}
\begin{equation}
|\Psi^{(\pm)}\rangle = \frac{1}{\sqrt{N!}}\left[\int^{\infty}_{-\infty}dz\left(\Psi_{\mp}\hat{a}_1^\dag - \Psi_{\pm}\hat{a}_2^\dag\right)\right]^N |0\rangle\\,
\end{equation}
with
\begin{eqnarray}
\Psi_{\pm}=\frac{\sqrt{NU}}{4}(1\pm|p_0|)\sech{\left(\frac{NU}{4}(1\pm|p_0|)z\right)},
\end{eqnarray}
\end{subequations}
and $|p_0| = \sqrt{\frac{1}{2\alpha}(1+\alpha-\frac{\Lambda}{2})}$.
By defining macroscopic superposition of states from Eqs.~\eqref{SCS_wf}, we can construct Schr\"odinger-cat states (SCS) from coupled solitons, cf.~\cite{cirac}:
\begin{equation}\label{SC}
|\Psi\rangle = C\left(|\Psi^{(+)}\rangle + |\Psi^{(-)}\rangle \right).
\end{equation}
Here, $C={[2(1 + X^N)]}^{-1/2}$ is a normalization factor, and $X=\frac{1-p_0^2}{2}\int_{-\infty}^{\infty}\frac{dx}{\cosh{[x]}+\cosh{[p_0x]}}\approx\left(1-p_0^2\right)\left(1-\alpha p_0^2\right)$ with the same $\alpha=0.21$.
It is noted that that SCS given in Eqs.~\eqref{SCS_wf} is not orthogonal to each other, but follows the following relation:
\begin{equation}\label{cat_size}
\epsilon=\langle \Psi^{(\pm)}|\Psi^{(\mp)} \rangle = X^N.
\end{equation}

\begin{figure}[t]
\includegraphics[width=8.4cm]{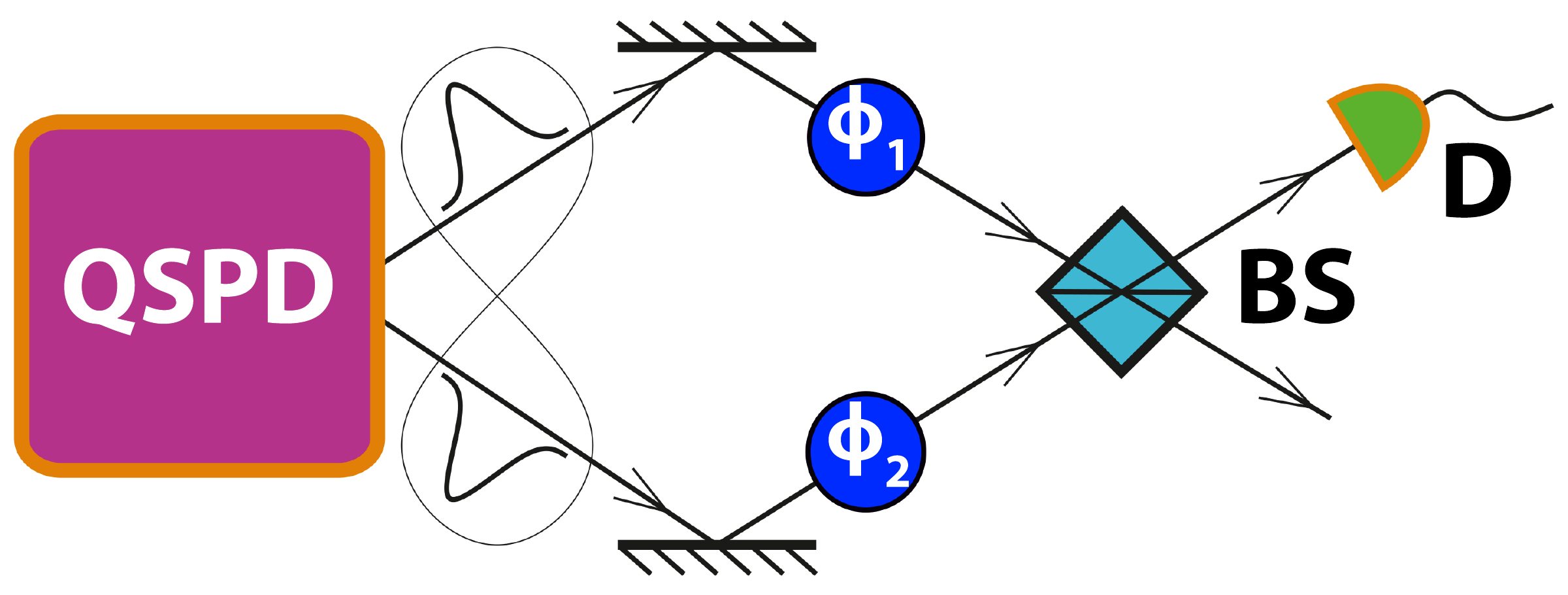}
\caption{The dependence of the ``cat size" $1/\epsilon$~\eqref{cat_size} on population imbalance $|p_0|$ for different numbers of particles $N$. One can see that ``cat size" tends to infinity when $|p_0|$ tends to 1. Also $1/\epsilon\approx0$ when $|p_0|\approx0$. Infinite ``cat size" corresponds to macroscopic SCS and one can take it as approximately N00N-state. Zero ``cat size" corresponds to microscopic SCS which means almost no entanglement.}
\label{fig:Cat_Size}
\end{figure}

Physically, the {\it size of the cat} can be defined by  $1/\epsilon$ (see Fig.~\ref{fig:Cat_Size}). For macroscopic SCS,  we ask  $\epsilon \ll 1$, which implies the maximally achievable cat size obtained with
$|p_0|\rightarrow1$ and $X\rightarrow0$.

\subsection{$N00N$-states}
The second set of solutions given in Eqs.~\eqref{sol2} presumes
\begin{subequations}\label{N00N_wf}
\begin{equation}
|\Phi^{(\pm)}\rangle = \frac{1}{\sqrt{N!}}\left[\int^{\infty}_{-\infty}dz\left(\Phi \hat{a}_{2,1}^\dag\right)\right]^N|0\rangle,
\end{equation}
with
\begin{equation}
\Phi = \frac{\sqrt{NU}}{2}\sech{\left(\frac{NU}{2}z\right)}.
\end{equation}
\end{subequations}
The superposition state  constructed from Eqs.~\eqref{N00N_wf} is:
\begin{equation}\label{N00N}
|\Phi\rangle = \frac{1}{\sqrt{2}}\left(|\Phi^{(+)}\rangle + e^{-i\theta_N} |\Phi^{(-)}\rangle \right),
\end{equation}
which clearly gives us a N00N-state of solitons.
Here, we also introduce $\theta_N = N\theta_0 = N \arccos\left(-\frac{\Lambda}{2(1-\alpha)}\right)$.
At the critical value of $\Lambda=\Lambda_{cr}=1.58$,  the SCS shown in Eq.~\eqref{SC} can be  transformed into $N00N$-state in Eq.~\eqref{N00N}, with a $\theta_0=\pi$ phase difference between two solitons.

\begin{figure}[t]
\includegraphics[width=8.4cm]{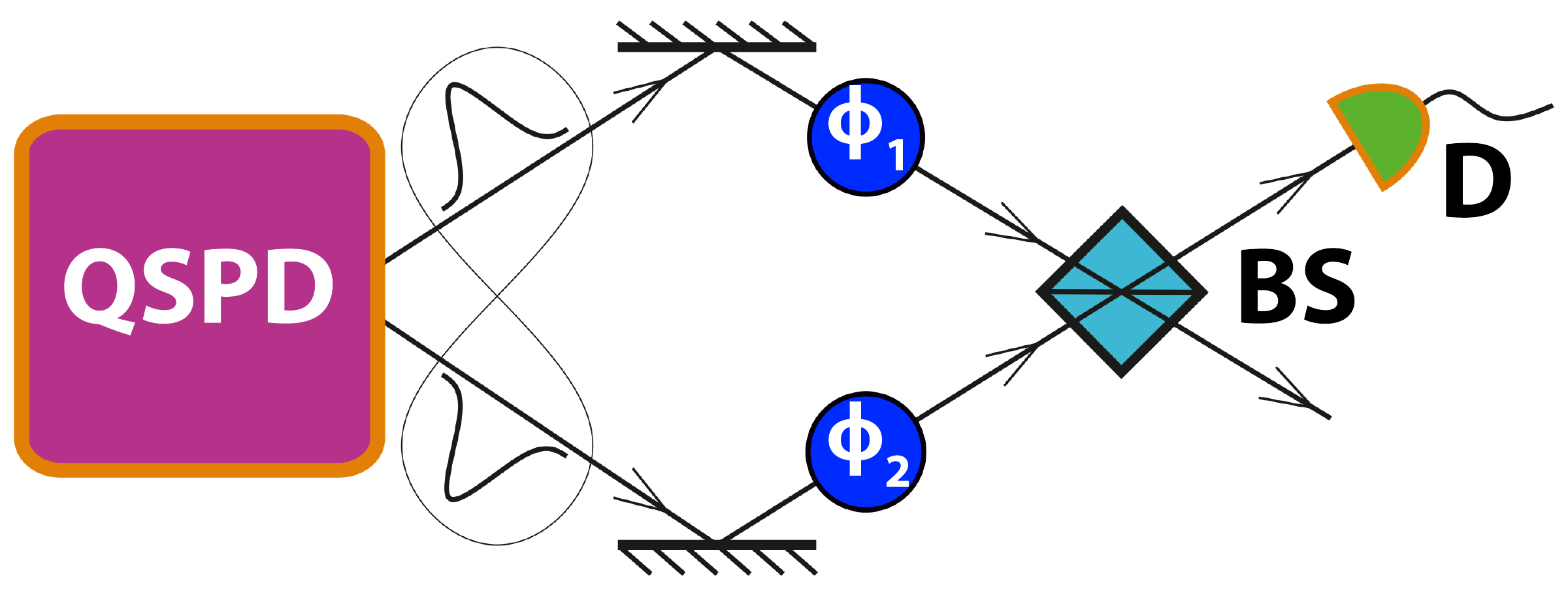}
\caption{Illustration of the precision measurement of the phase shift, based on a  Mach-Zehnder interferometer (MZI). Here, QSPD denotes a quantum state preparation device, $\phi_1$ and $\phi_2$ are two resulting phases accumulated at the arms of interferometer, $BS$ is a beam-splitter, and $D$ is a parity detector that runs in the particle counting regime.}
\label{fig:MZI}
\end{figure}

\section{Quantum measurements with superposition states}
In this section we propose a precision measurement experiment with SCS and $N00N$-state. A Mach-Zehnder interferometer (MZI) is illustrated in Fig.\ref{fig:MZI}. The device coined as a quantum state preparation device (QSPD) represents the medium with two coupled BECs producing entangled soliton states (may be the superposition state, SCS or $N00N$-state) into the input of a MZI. The measured parameter is a linear phase shift $\phi= \phi_2-\phi_1$ accumulated in the arms of MZI.

The sensitivity of the phase parameter $\phi$ for scheme in Fig. \ref{fig:MZI}
is determined by (cf.\cite{Helstrom})
\begin{equation}
\langle\left(\Delta\phi\right)^2\rangle = \frac{\langle\left(\Delta\hat{P}\right)^2\rangle}{\left|\frac{\partial\langle\hat{P}\rangle}{\partial\phi}\right|^2},
\label{estimation}
\end{equation}
where, $\hat{P}$ is a Hermitian operator suitable for the measurement of  the phase $\phi$. We propose to use parity detection procedure  with an operator taken for the second mode:
$\hat{P}\equiv\hat{P}_{\hat{a}_2}=\exp\left[i\pi\int_{-\infty}^{\infty}\hat{a}_2^{\dag}\hat{a}_2dz\right]$.
In order,  for parity measurement shown in Fig. \ref{fig:MZI}, two matter waves after phase-shifting operations, are superimposed in the beam-splitter, and then one of the detectors counts even or odd number in particles, cf. [34].

To describe the parity measurement, one may introduce spin operators as follows:
\begin{subequations}\label{Spin}
\begin{equation}
\hat{S}_0 = \frac{1}{2}\int_{-\infty}^{\infty}\left(\hat{a}^{\dag}_1\hat{a}_1+\hat{a}^{\dag}_2\hat{a}_2\right)dz, 
\end{equation}
\begin{equation}
\hat{S}_1 = \frac{1}{2}\int_{-\infty}^{\infty}\left(\hat{a}^{\dag}_1\hat{a}_1-\hat{a}^{\dag}_2\hat{a}_2\right)dz, 
\end{equation}
\begin{equation}
\hat{S}_2 = \frac{1}{2}\int_{-\infty}^{\infty}\left(\hat{a}^{\dag}_1\hat{a}_2+\hat{a}^{\dag}_2\hat{a}_1\right)dz, 
\end{equation}
\begin{equation}
\hat{S}_3 = \frac{i}{2}\int_{-\infty}^{\infty}\left(\hat{a}^{\dag}_2\hat{a}_1-\hat{a}^{\dag}_1\hat{a}_2\right)dz.
\end{equation}
\end{subequations}
These operators obey $SU(2)$ algebra and obey to commutation relations: $[\hat{S}_i, \hat{S}_j] = i\epsilon_{ijk}\hat{S}_k$, with  $i,j,k=1,2,3$.
With $\hat{S}_j$  operators, we can define unitary operators for the transformations of  quantum state in the beam-splitter and phase shift, i.e.,  $\hat{U}_{BS} = \exp\left[i\frac{\pi}{2}\hat{S}_2\right]$ and $\hat{U}_{PS} = \exp\left[-i\phi\hat{S}_1\right]$, respectively.
Then, the action of a MZI on initial quantum state can be described by MZI-operator, i.e., $\hat{U}_{MZI} = \hat{U}_{BS}\hat{U}_{PS} = \exp\left[i\frac{\pi}{2}\hat{S}_2\right]\exp\left[-i\phi\hat{S}_1\right]$. The parity operator $\hat{P}_{a_2}$ in this formalism has a form:
\begin{equation}
\hat{P}_{\hat{a}_2} \equiv \exp\left[i\pi(\hat{S}_0 - \hat{S}_1)\right].
\label{parity}
\end{equation}
Thus, for the scheme shown in Fig. \ref{fig:MZI},  the resulting expectation value with parity operator $\hat{P}_{a_2}$ can be calculated as
\begin{equation}
\langle \hat{P}_{\hat{a}_2}\rangle = \langle \hat{U}_{MZI}^{\dag}\hat{P}_{a_2}\hat{U}_{MZI}\rangle = \langle e^{i\pi\hat{S}_0}e^{i\phi\hat{S}_1}e^{i\pi \hat{S_3}}e^{-i\phi\hat{S}_1}\rangle.
\label{exp_parity}
\end{equation}

It is also more convenient to use an angular momentum state representation instead of particle number representation.
Here, we consider the substitution by $|N_1,N_2\rangle \rightarrow |j,m\rangle$,  where $N_1$, $N_2$ are numbers of particles in the first and the second wells.
The quantum numbers for angular momenta  $j$ and $m$ are introduced as $j = N/2$ and $m =(N_1-N_2)/2$, respectively.
The states $|j,m\rangle$ are eigenstates of the spin operators $\hat{S}_{0,1}$ with the  conditions $\hat{S}_1|j,m\rangle = m|j,m\rangle$; $\hat{S}_0|j,m\rangle = j|j,m\rangle$and $\exp\left[i\pi S_3\right]|N_1,N_2\rangle = \exp\left[i\pi N_1\right]|N_2,N_1\rangle$.

\begin{figure}[h!]
\includegraphics[width=3.2in]{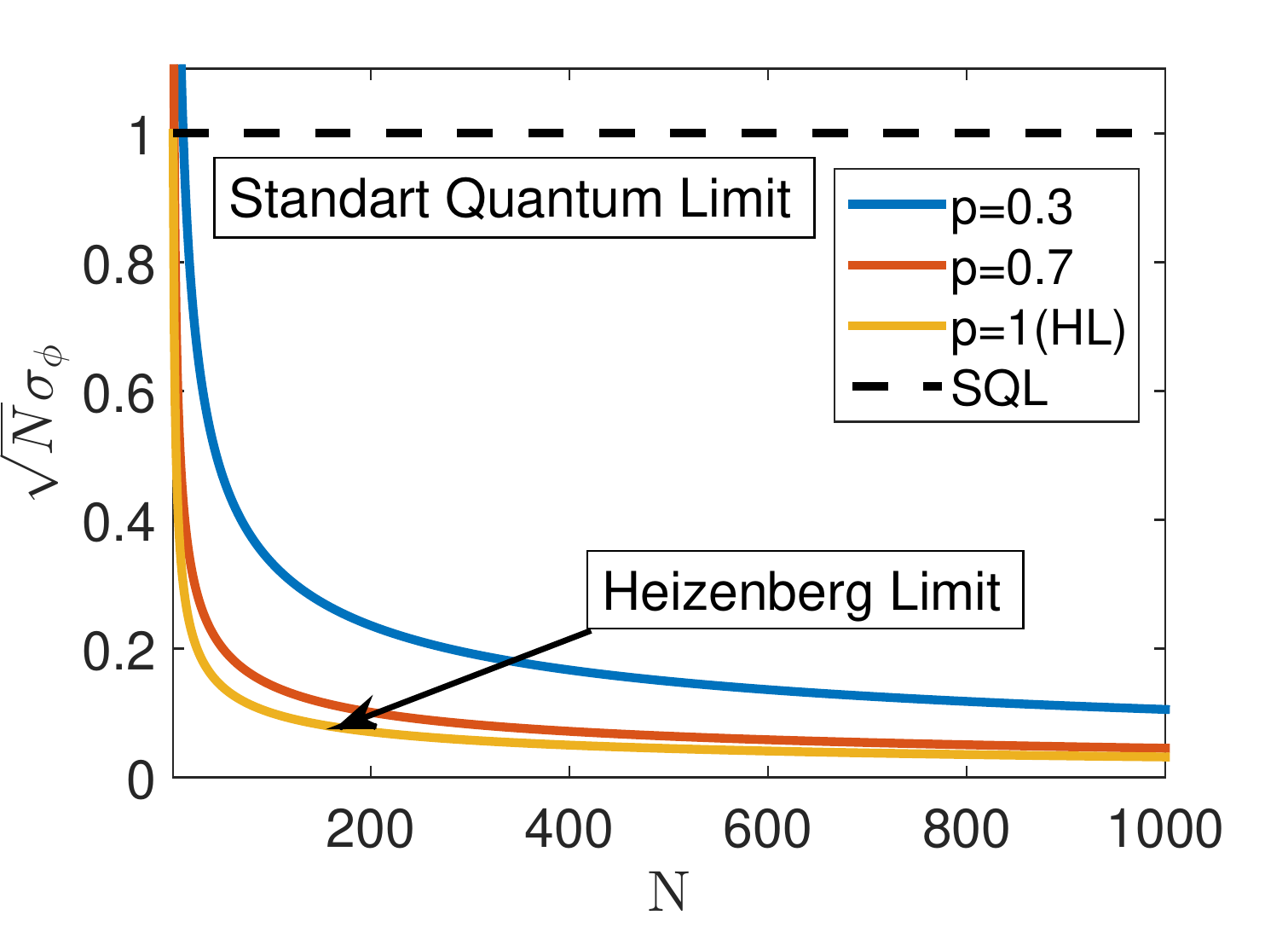}
\caption{Reduced phase uncertainty $\sqrt{N}\sigma_{\phi}$ against total particle number $N$, for an initial SCS used  in the measurement procedure. The value $\sqrt{N}\sigma_{\phi}=1$ corresponds to SQL limit.}
\label{fig:error_N}
\end{figure}

In terms of the angular momentum we can rewrite SCS in Eq.~\eqref{SC} and $N00N$state Eq.~\eqref{N00N}  as
\begin{subequations}
\begin{eqnarray}
&& |\Psi\rangle = C\left(|j,-j|p_0|\rangle + |j,j|p_0|\rangle\right),\\
&& |\Phi\rangle = \frac{1}{\sqrt{2}}\left(|j,-j\rangle + e^{-i\theta_{N}}|j,j\rangle\right).
\end{eqnarray}
\end{subequations}
Then, for measured parity operator, the resulting average value $\langle \hat{P}_{a_2}\rangle$ for initial SCS and $N00N$-states, respectively, have the form:
\begin{subequations}
\begin{eqnarray}
&& \langle\Psi|\hat{P}_{\hat{a}_2}|\Psi\rangle = (-1)^N\cos\left[\left(\phi-\frac{\pi}{2}\right)N|p_0|\right],\\
&& \langle\Phi|\hat{P}_{\hat{a}_2}|\Phi\rangle =
 \begin{cases}
   (-1)^{\frac{N}{2}}\cos\left[\phi N + \theta_N\right]; &\text{$N$ is even}\\
   (-1)^{\frac{N+1}{2}}\sin\left[\phi N + \theta_N\right];  &\text{$N$ is odd}
 \end{cases}\nonumber\\
\end{eqnarray}
\end{subequations}
with the  variation $\langle(\Delta \hat{P}_{\hat{a}_2})^2\rangle$:
\begin{subequations}
\begin{eqnarray}
&& \langle\Psi|(\Delta \hat{P}_{\hat{a}_2})^2|\Psi\rangle = \sin^2\left[\left(\phi-\frac{\pi}{2}\right)N|p_0|\right],\\
&& \langle\Phi|(\Delta \hat{P}_{\hat{a}_2})^2|\Phi\rangle =
 \begin{cases}
   \sin^2\left[\phi N + \theta_N\right]; &\text{$N$ is even}\\
   \cos^2\left[\phi N + \theta_N\right];&\text{$N$ is odd}
 \end{cases}
\end{eqnarray}
\end{subequations}
From the results above, we can see that quantum interference effects arise in the  parity measurement scheme,  depending on even or odd particle numbers $N$.
As for the sensitivity of interferometer, from Eq.~\eqref{estimation}  we immediately obtain
\begin{subequations}
\begin{eqnarray}
&& \langle\Psi|(\Delta\phi)^2|\Psi\rangle = \frac{1}{N^2|p_0|^2},\\
&& \langle\Phi|(\Delta\phi)^2|\Phi\rangle = \frac{1}{N^2}.
\end{eqnarray}
\end{subequations}
One can see that the Heisenberg limit is achieved for a maximally entangled $N00N$-state and a precision for SCS has an extra $1/|p_0|^2$ factor. In Fig.~\ref{fig:error_N}, we plot the  normalized error in phase measurement $\sigma_{\phi}=\sqrt{\langle(\Delta\phi)^2\rangle}$ as a function of particle number $N$ for SCS. The value  $\sigma_{\phi}= N^{-1/2}$  characterizes SQL of phase measurement with classical states, which can  be achieved without QSPD. One can see that accuracy of a measurement tends to the Heisenberg limit as the cat size grows and saturates with $|p_0|=1$ at the input (the yellow curve in Fig.~\ref{fig:error_N}). On the contrary, a microscopic SCS obtained with $|p_0|\rightarrow0$ is not suitable to perform the measurements.

\section{Measurements beyond the Heisenberg scaling}
The accuracy of measurement can be improved even more by using parameters with nonlinear particle number dependence. 
In the framework of nonlinear interferometry, for arbitrary $\Theta$-parameter measurement procedure uses transformation $|\Psi\rangle_\Theta = \exp(i\Theta G)|\Psi\rangle$ for input state $|\Psi\rangle$, where $G$ is the generator of transformation that describers nonlinear phase dependence, cf. ~\cite{roy,boixo, Luis, Mundo}. 

 In general case for $G=N^k$ limiting sensitivity of the $\Theta$-parameter measurement for nonlinear interferometer  in bounded by the value $\sigma_{\Theta} \simeq 1/N^k$, which corresponds to so-called super-Heisenberg limit for phase measurement in quantum metrology, cf. \cite{napolitano}. 

Let us examine the measurement of the parameter $\Theta=\frac{\Lambda}{N^2}=\frac{U^2}{16|\kappa|}$ instead of phase shift $\phi$ by using $N00N$-state with the initial phase difference
\begin{equation}
\theta_N = N \arccos\left(-\frac{\Theta N^2}{2(1-\alpha)}\right),
\end{equation}
see Eqs.~(\ref{N00N},~\ref{estimation}).
\begin{figure}[t]
\includegraphics[width=8.4cm]{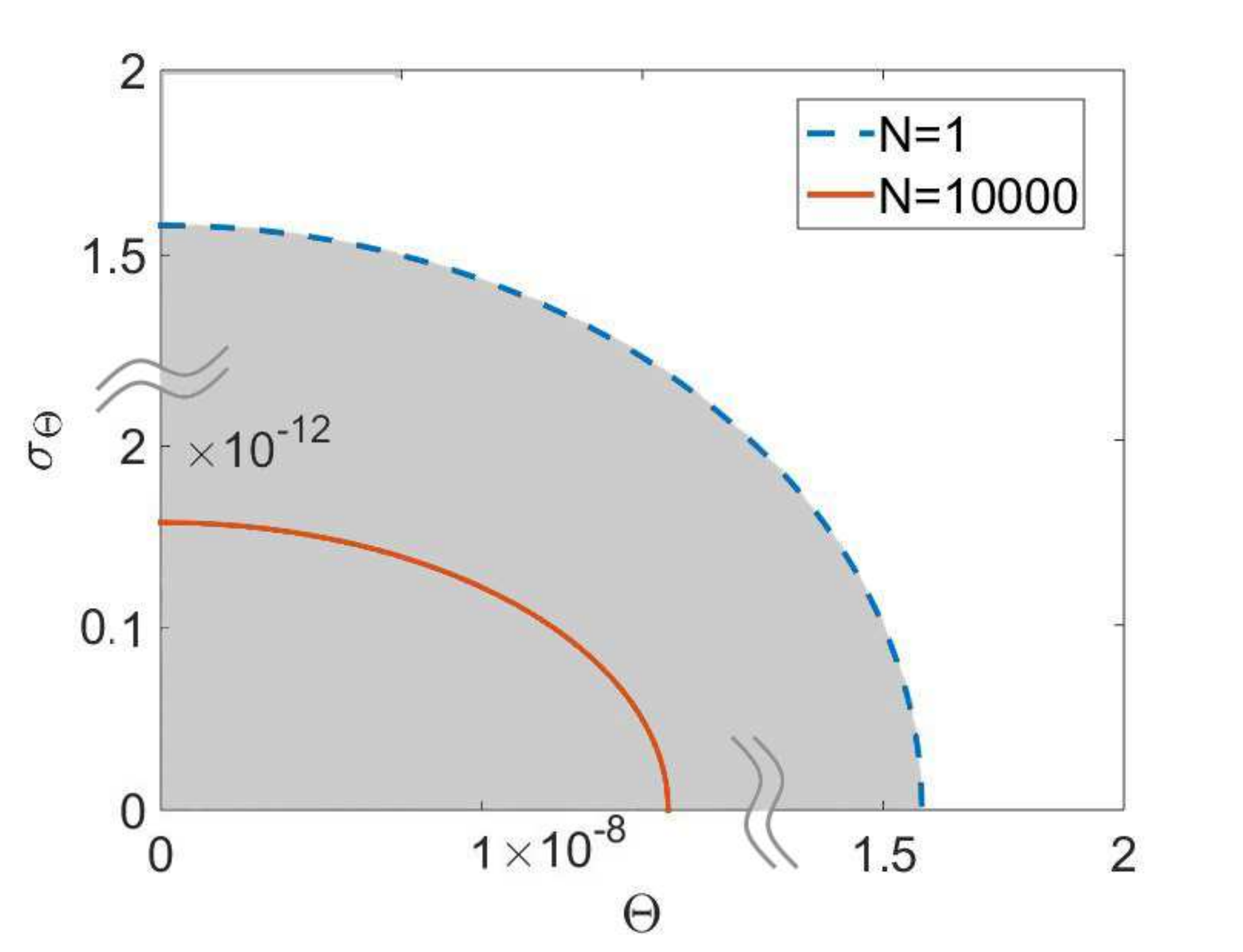}
\caption{The dependence of $\sigma_{\Theta}$ on $\Theta$ demonstrating second-order like phase transition from the state possessing non-zero $\sigma_{\Theta}$ beyond the linear Heisenberg limit (gray area) to the nonapplicable for such measurements state.}
\label{FIG:L0}
\end{figure}

For a sufficiently small $\Theta$, we can apply the Taylor expansion
\begin{equation}
\theta_N = \frac{\pi}{2}N + \frac{N^3}{2(1-\alpha)}\Theta + O(\Theta^3),
\label{Taylor}
\end{equation}
which is valid as long as we take account only linear dependence on $\Theta$.
By setting  $\phi=0$ to neglect unimportant phase shift, we have for the $N00N$-state at the input of the MZI:

\begin{subequations}
\begin{eqnarray}
&&\langle\Phi|\hat{P}_{\hat{a}_2}|\Phi\rangle =
 \begin{cases}
   (-1)^{\frac{N}{2}}\cos\left[\theta_N\right]; &\text{$N$ is even}\\
   (-1)^{\frac{N+1}{2}}\sin\left[\theta_N\right];  &\text{$N$ is odd}
 \end{cases}\\
&&\langle\Phi|(\Delta \hat{P}_{\hat{a}_2})^2|\Phi\rangle =
 \begin{cases}
   \sin^2\left[\theta_N\right]; &\text{$N$ is even}\\
   \cos^2\left[\theta_N\right];&\text{$N$ is odd}
 \end{cases}
\end{eqnarray}
\end{subequations}
for the  average value of $\hat{P}_{a_2}$ and the corresponding variance, respectively.
The resulting sensitivity of  $\Theta$ can be found to be:
\begin{equation}
\langle\left(\Delta\Theta\right)^2\rangle =\frac{4(1-\alpha)^2}{N^6}.
\label{disp_L}
\end{equation}


From Eq. (\ref{disp_L}), the error in $\Theta$ measure is $\sigma_{\Theta}=\sqrt{\langle(\Delta\Theta)^2\rangle}\sim N^{-3}$ that looks quite promising for improvement of  measurement sensitivity  currently achieved with atomic condensates, cf.  \cite{napolitano, Pezze, morsch}.    

In Fig.~\ref{FIG:L0}, we show the dependence of $\sigma_{\Theta}$ as a function of measured $\Theta$-parameter,  for different particle numbers $N$.
The dashed-curve in Fig. \ref{FIG:L0} corresponds to the limiting measurements with one particle.
The dependences in Fig. \ref{FIG:L0} demonstrate a second-order like continuous quantum phase transition from  the state possessing non-zero $\sigma_{\Theta}$ beyond the linear Heisenberg scaling  (through $NOON$-state) to the nonapplicable for such measurements state.
The shadowed region in Fig. \ref{FIG:L0} reveals the capacity  for measurements with particle number $N > 1$.

\section{Conclusion}
In summary, by adopting  quantum field theory approach to the problem of bright matter wave  soliton formation in  weakly coupled double-well potentials, we reveal the ground states in the
 Schr\"odinger-cat  superposition (SCS) states and maximally path entangled $N00N$-states.
With variational method, we derive the equation of motions for SCS and $N00N$-states.
Then,  within the Mach-Zehnder interferometer we examine quantum phase measurement with these superposition states, in order to have the accuracy beyond the standard quantum limit and the linear Heisenberg limit.
We perform the $\hat{P}_{\hat{a}_2}$ operator measurements by applying parity measurement procedure.
Heisenberg-limited phase shift measurements are demonstrated to be saturated for maximally path entangled states containing $N$ particles.  Vital combination of condensate parameters $\Theta=\frac{U^2}{16|\kappa|}$ is shown to surpass the linear Heisenberg limit in the framework of nonlinear metrology approach, with a  scaling  proportional to $N^{-3}$. This results applied for atomic $N00N$-states represent a promising tool for atomic clocks and atomic gyroscopes~\cite{Pezze, morsch}.

Notably,  decoherence effects play an important role for the schemes operating with SCS and/or $N00N$-states (cf.\cite{Haroche}). 
Although, at present parity measurement represents experimentally non-trivial task requiring high efficiency  single particle counting detectors it is absolutely  necessary to achieve Heisenberg scaling with phase measurement in our scheme, cf. (cf.\cite{gerry}).


 From the practical point of view it is more important  to discuss characteristic time scales when superposition states and more generally -- two component macroscopic condensates can be useful for quantum operations. Contrary to standard (single particle) qubits, as it is shown in Ref.~\cite{Tim}, the required time of  gate operation in condensates  to produce entanglement is inversely proportional to the particle number $N$. This enhancement is achieved due to bosonic stimulation effect and implies fast quantum gate operation. Obviously, decoherence effects occurring with condensate macroscopic states should appear at longer time. In this case there are more physical systems which are capable for fast operations. 
 
Our results provide useful information not only for atomic optics, but also solid state physics technology. 
 In particular, exciton-polariton bright solitons in high-Q semiconductor microcavities  represent one of them \cite{sich}. The lifetime of solitons is several tens of picoseconds that is large enough in comparison with possible quantum operation. Moreover, recently proposed by Y. Sun  {\it et al.} in Ref. \cite{Sun} proposed for semiconductor microstructures with few hundred picoseconds lifetime for low branch exciton polaritons which enables to use such systems for mentioned purposes. In other words, long-lived exciton polariton condensates~\cite{Sun, sich}), can be a new platform to design maximally entangled states.

\section*{Acknowledgment}
We acknowledge the financial support from RFBR, Grants No. 15-52-52001 and No. 14-02-97503.




\begin{thebibliography}{}

\bibitem{wiseman}
H. M. Wiseman and G. J. Milburn, {\it Quantum Measurement and Control,} (Cambridge University Press, 2010).

\bibitem{wieman}
C. E. Wieman, D. E. Pritchard, and David J. Wineland, ``Atom cooling, trapping, and quantum manipulation,'' Rev. Mod. Phys. {\bf 71}, S253 (1999).

\bibitem{dowling15}
R. Demkowicz-Dobrzanski, M. Jarzyna, J. Kolodynski, "Quantum limits in optical interferometry", 
Progress in Optics {\bf 60}, 345 (2015); J. P. Dowling and K. P. Seshadreesan, ``Quantum Optical Technologies for Metrology, Sensing, and Imaging,'' J. of Lightwave Tech., {\bf 33}, 2359 (2015).

\bibitem{giovannetti}
V. Giovannetti, S. Lloyd, and L. Maccone, ``Quantum-Enhanced Measurements: Beating the Standard Quantum Limit,'' Science {\bf 306}, 1330, (2004); V. Giovannetti, S. Lloyd, and L. Maccone, ``Advances in quantum metrology,''
Nature Photon. 5, 222 (2011).

\bibitem{roy}
S. M. Roy and S. L. Braunstein, ``Exponentially Enhanced Quantum Metrology,'' Phys. Rev. Lett. {\bf 100}, 220501 (2008).

\bibitem{boixo}
S. Boixo, Steven T. Flammia, C. M. Caves, and J.M Geremia, ``
Generalized Limits for Single-Parameter Quantum Estimation,''  Phys. Rev. Lett. {\bf 98}, 090401 (2007); S. Boixo, A. Datta, M. J. Davis, S. T. Flammia, A. Shaji, and C. M. Caves, ``Quantum Metrology: Dynamics versus Entanglement,'' Phys. Rev. Lett, {\bf 101}, 040403 (2008).

\bibitem{Caves81}
C. M. Caves, ``Quantum-mechanical noise in an interferometer,'' Phys. Rev. D {\bf 23}, 1693 (1981).

\bibitem{yurke}
B. Yurke, S. L. McCall, and J. R. Klauder, ``SU(2) and SU(1,1) interferometers,'' Phys. Rev. A {\bf 33}, 4033 (1986); P. Grangier, R. E. Slusher, B. Yurke, and A. LaPorta, ``Squeezed-light–enhanced polarization interferometer,'' Phys. Rev. Lett. {\bf 59}, 2153 (1987).

\bibitem{gustavson}
T. L. Gustavson, P. Bouyer, and M. A. Kasevich, ``Precision Rotation Measurements with an Atom Interferometer Gyroscope,'' Phys. Rev. Lett. 78, 2046 (1997).

\bibitem{gross}
C. Gross, T. Zibold, E. Nickolas, J. Est\`{e}ve, M.K. Oberthaler, ``Nonlinear atom interferometer surpasses classical precision limit,'' Nature {\bf 464}, 1165 (2010).

\bibitem{dowling98}
J. P. Dowling, ``Correlated input-port, matter-wave interferometer: Quantum-noise limits to the atom-laser gyroscope,''  Phys. Rev. A {\bf 57}, 4736 (1998).

\bibitem{winland}
D. J. Wineland, J. J. Bollinger, W. M. Itano, F. L. Moore, and D. J. Heinzen, ``Spin squeezing and reduced quantum noise in spectroscopy,'' Phys. Rev. A {\bf 46}, R6797 (1992); J. J. Bollinger, W. M. Itano, D. J. Wineland, and D. J. Heinzen, ``Optimal frequency measurements with maximally correlated states,'' Phys. Rev. A {\bf 54}, R4649 (1996).

\bibitem{boto}
A. N. Boto, P. Kok, D. S. Abrams, S. L. Braunstein, C. P. Williams, and J. P. Dowling,
``Quantum Interferometric Optical Lithography: Exploiting Entanglement to Beat the Diffraction Limit,'' Phys. Rev. Lett. {\bf 85}, 2733 (2000); P. Kok, S.L. Braunstein, and J. P. Dowling, ``Quantum lithography, entanglement and Heisenberg-limited parameter estimation,'' J. Opt. B: Quantum Semiclass. Opt. {\bf 6}, S811 (2004).


\bibitem{dowling08}
J. P. Dowling, ``Quantum optical metrology – the lowdown on high-N00N states,'' Contem. Phys. {\bf 49}, 125 (2008).

\bibitem{pezze}
L. Pezze and A. Smerzi, ``Mach-Zehnder Interferometry at the Heisenberg Limit with Coherent and Squeezed-Vacuum Light,''
Phys. Rev. Lett. {\bf 100}, 073601 (2008).

\bibitem{afek}
I. Afek, O. Ambar, and Y. Silberberg, ``High-NOON States by Mixing Quantum and Classical Light,'' Science {\bf 328}, 879 (2010).

\bibitem{rozema}
L. A. Rozema, J. D. Bateman, D. H. Mahler, R. Okamoto, A. Feizpour, A. Hayat, and A. M. Steinberg, ``Scalable Spatial Superresolution Using Entangled Photons,'' Phys. Rev. Lett. {\bf 112}, 223602 (2014).

\bibitem{Luis}
J. Beltran and A. Luis, ``Breaking the Heisenberg limit with inefficient detectors,'' Phys. Rev. A {\bf 72}, 045801 (2005).

\bibitem{napolitano}
M. Napolitano, M. Koschorreck, B. Dubost, N. Behbood, R. J. Sewell, and M. W. Mitchell, ``Interaction-based quantum metrology showing scaling beyond the Heisenberg limit,'' Nature {\bf 471}, 486 (2011).


\bibitem{Mundo}
D. Maldonado-Mundo and A. Luis, "Metrological resolution and minimum uncertainty states in linear and nonlinear signal detection schemes", Phys.  Rev. A {\bf 80}, 063811 (2009).


\bibitem{pethick}
C. J. Pethick and H. Smith, {\it Bose-Einstein Condensation in Dilute Gases,} (Cambridge University Press, 2008).


\bibitem{cirac}
J. I. Cirac, M. Lewenstein, K. M$\o$lmer, and P. Zoller, ``Quantum superposition states of Bose-Einstein condensates,'' Phys. Rev. A {\bf 57}, 1208 (1998).

\bibitem{Soren}
A. Sorensen, L.-M. Duan, J. I. Cirac and P. Zoller, ``Many-particle entanglement with Bose-Einstein condensates,'' Nature, {\bf 409}, 63 (2001).

\bibitem{Fu}
L. Fu and J. Liu, ``Quantum entanglement manifestation of transition to nonlinear self-trapping for Bose-Einstein condensates in a symmetric double well,'' Phys. Rev.  A {\bf 74}, 063614 (2006).

\bibitem{Maz}
G. Mazzarella, L. Salasnich, A. Parola, and F. Toigo, ``Coherence and entanglement in the ground state of a bosonic Josephson junction: From macroscopic Schr\"odinger cat states to separable Fock states,'' Phys. Rev. A {\bf 83}, 053607 (2011).

\bibitem{He}
Q. Y. He, P. D. Drummond, M. K. Olsen, and M. D. Reid, ``Einstein-Podolsky-Rosen entanglement and steering in two-well Bose-Einstein-condensate ground states,'' Phys. Rev. A {\bf 86}, 023626 (2012).

\bibitem{Pezze}
L. Pezze, A. Smerzi, M. K. Oberthaler,  R. Schmied, and P. Treutlein,
``Quantum metrology with nonclassical states of atomic ensembles,''  arXiv:1609.01609 (2016).

\bibitem{morsch}
O. Morsch and M. Oberthaler, ``Dynamics of Bose-Einstein condensates in optical lattices,'' Rev. Mod. Phys. {\bf 78}, 179 (2006).

\bibitem{Hof}
C. Schneider, K. Winkler, M. D. Fraser, M. Kamp, Y. Yamamoto, E. A. Ostrovskaya, and S. Hofling,
``Exciton-polariton trapping and potential landscape engineering,''
Rep. Prog. Phys. {\bf 80}, 016503 (2017).



\bibitem{Alod}
A. P. Alodjants and S. M. Arakelian,
``Quantum nondemolition measurements of the phase and polarization Stokes parameters of optical fields,''
Zh. Eksp. Teor. Fiz. {\bf 13}, 1235 (1998).


\bibitem{raghavan}
S. Raghavan and G. P. Agrawal,
``Switching and self-trapping dynamics of Bose-Einstein solitons,'' J. Mod. Opt. {\bf 47}, 1155 (2000).

\bibitem{strecker}
K. E. Strecker, G. B. Partridge, A. G. Truscott, and R. G. Hulet, ``Formation and propagation of matter-wave soliton trains,'' Nature {\bf 417}, 150 (2002); L. Khaykovich, F. Schreck, G. Ferrari, T. Bourdel, J. Cubizolles, L. D. Carr, Y. Castin, and C. Salomon,
``Formation of a Matter-Wave Bright Soliton,'' Science {\bf 296}, 1290 (2002).

\bibitem{Helstrom}
C.W. Helstrom, {\it Quantum Detection and Estimation Theory, Mathematics in
Science and Engineering,}  (Academic Press, 1976).

\bibitem{Haroche}
S. Haroche and J.-M. Raimond, {\it Exploring the Quantum}  (Oxford University Press, 2006).

\bibitem{gerry}
C. C. Gerry, A. Benmoussa, and R. A. Campos, ``Parity measurements, Heisenberg-limited phase estimation, and beyond,''
J. Mod. Opt. {\bf 54}, 2177 (2007).

\bibitem{Tim}
T. Byrnes, Kai Wen, and Y. Yamamoto, "Macroscopic quantum computation using Bose-Einstein condensates," Phys. Rev. A {\bf 85},  040306(R) (2012).

\bibitem{sich}
M. Sich, D. N. Krizhanovskii, M. S. Skolnick, A. V. Gorbach, R. Hartley, D. V. Skryabin, E. A. Cerda-Méndez, K. Biermann, R. Hey, and P. V. Santos,
``Observation of bright polariton solitons in a semiconductor microcavity,''
Nat. Photon. {\bf 6}, 50 (2012).

\bibitem{Sun}
Y. Sun, P. Wen, Y. Yoon,  G. Liu, M. Steger, L. N. Pfeiffer, K. West, D. W. Snoke, and K. A. Nelson,
``Bose-Einstein Condensation of Long-Lifetime Polaritons in Thermal Equilibrium,''
Phys. Rev. Lett., {\bf 118}, 016602 (2017).




\end{thebibliography}
\end{document}